\documentclass[psfig]{mn}  
\usepackage{graphicx} % standard LaTeX graphics tool
\begin{document}
\title[Cores of Dark Matter Halos Correlate with Stellar Scale Lengths]
{Cores of Dark Matter Halos Correlate with Stellar Scale Lengths}
\author [F. Donato, G. Gentile and P. Salucci]
{Fiorenza Donato $^{1}$\thanks{E-mail:donato@to.infn.it},
Gianfranco Gentile$^{2}$\thanks{E-mail:gentile@sissa.it},
and Paolo Salucci$^{2}$\thanks{E-mail:salucci@sissa.it}\\
$^{1}$ Universit\`a degli Studi di Torino
and INFN, via Giuria 1, 10125 Torino, Italy\\
$^{2}$ SISSA, via Beirut, 2-4 34014 Trieste, Italy}

%\authorrunning{Donato Salucci}
%\institute{SISSA, Via Beirut, 2-4
%I-34014 Trieste, Italy, {\it salucci@sissa.it}}
\maketitle % typesets the title of the contribution
\begin{abstract}
We investigate in detail the mass distribution obtained by means of high
resolution rotation curves of 25 galaxies of different morphological types. 
The dark matter contribution to the circular rotation velocity is 
well-described by resorting to a dark component whose density shows an inner
core, i.e.  a central  constant density region. We find a very strong
correlation between the core radius size $R_C$ and the stellar  exponential
scale length $R_D$: 
$R_C \simeq  13 \  (\frac {R_D} {5\ {\rm kpc}})^{1.05} \ {\rm kpc} $,  and
between  $R_C$ and the galaxy dynamical mass at this distance, $M_{dyn}(R_C)$.
These relationships would not be expected if the core radii  were
the product of an incorrect decomposition procedure, or the biased result of
wrong or misunderstood observational data. The very strong correlation between
the dark and luminous scale lengths found here seems to hold also for different 
Hubble types and opens new scenarios for the nature of the dark matter 
in galaxies.
\end{abstract}


\section{Introduction}
\label{sec:introduction}
In spite of a large amount of knowledge gathered over the past 20 years on
the phenomenon of Dark Matter (DM) in Galaxies, it is only recently that the
attention has been focused on its radial density profile. Recall 
that, according to high-resolution N-body simulations in the $\Lambda$-CDM
framework, the dark halo density, characterized by the virial halo mass as
free parameter, is described by an inner power--law cusp (but see Ricotti 2003):
\begin{equation}
\rho_{\rm NFW}(r) = \frac{\rho_s}{(r/r_s)(1+r/r_s)^2}
\label{eq:rho_nfw}
\end{equation}
where $r_s$ and $\rho_s$ are related parameters (Navarro,
Frenk and White, 1996b, hereafter NFW). The DM contribution to the circular
velocity can be written as:
\begin{equation}
V_{\rm NFW}^2(r)= V_{vir}^2 \frac{c}{A(c)} \frac {A(x)}{x}
\end{equation}
where $x \equiv   c r/R_{vir}$, $c \equiv R_{vir}/r_s$ is the 
concentration parameter, and $A(x)\equiv \ln (1+x) - x/(1+x)$.
We also have that $V^2_{vir}= G M_{vir}/ R_{vir}$, 
$\rho_s \simeq 101/3 \ \rho_{crit} \ c^3 \ / (\ln(1+c)-c/(1+c))$ 
and $M_{vir}  = \frac{4 \pi}{3} \delta_{vir} \rho_{mean} R^3_{vir}$.
\footnote {$\rho_{crit}$ is the critical density of the universe, 
$\delta_{vir} \simeq 337$ is the virial overdensity and
$\rho_{mean}$ is the mean universal density at the galaxy's redshift. 
The values for the numerical constants are valid for a $\Lambda$-CDM 
cosmogony with $\Omega_0$=0.3 and $\Omega_\Lambda$=0.7.}

A number of reliable mass modellings, obtained for spirals and low surface
brightness galaxies (LSBs) has supported the early claim (Moore, 1994) that
dark halos around disk  galaxies have a central density distribution much
shallower than the NFW one (e.g. Gentile et al., 2004 and references therein;
Weldrake, de Blok \& Walter, 2003; de Blok \& Bosma, 2002). The first studies
on elliptical galaxies seem to indicate that also  these objects share the same
phenomenon (Gerhard et al., 2001;  Borriello et al., 2003 ).   

 This  ``galaxy by galaxy'' comparison between the predicted $\Lambda$-CDM
density distribution and those actually detected for the dark halos around disk
galaxies has been the main goal of several published works. However, although
the study of the discrepancy between data and theoretical predictions remains
necessary, we believe that time has come that we investigate {\it per se} the
distribution of DM around galaxies, independently of the existing cosmological
implications. In fact, the hot debate on possible falsification of
$\Lambda$-CDM on galactic scales has carried the research off from the original
topic of understanding the {\it DM phenomenon} in virialized systems. Let us
stress that a {\it direct} knowledge on the presence, nature and interaction
with baryons of the  dark galactic  component  is still very rough and limited,
unlike the complex and refined scenario that theory and simulations have put
forward.

In the decomposition of rotation curves (RC's), the circular velocity  $V(r)$
is best-fit  modelled as the sum of a baryonic component that  includes a stellar
and a gaseous disk, and a spherical dark halo:
\begin{equation}
V^2(r)=V_D^2(r)+V^2_{gas}(r)+V^2_H(r)
\end{equation}
 where the labels D, gas and H refers to the corresponding components.  It is
common to represent the dark halo  contribution to the circular velocity by
means  of the   2-parameter function describing the circular velocity of a
pseudo-isothermal (PI) halo:
\begin{equation}
V^2_{H,PI}(r) =  4\pi G \rho_0 R_C^2 \left( 1 - \frac{R_{C}}{r} 
{\rm arctan} \frac{r}{R_{C}} \right).
\label{eq:pi}
\end{equation}
The central density $\rho_0$ and the halo {\it density core} 
radius $R_C$ are  free parameters to be tuned to fit $V^2(r)$. 

Persic, Salucci \& Stel (1996, PSS hereafter)  introduced,  
for the  dark halo velocity contribution, an equally  simple function,
\begin{equation}
V^2_{H, URC}(r)=(1-\beta) V_{opt}^2 r^2 (1+a^2)/(r^2+a^2 R_{opt}^2)
\label{eq:urc}
\end{equation}
to fit, once   added to the  disk contribution,  the  Universal Rotation Curve
(URC) of Spirals,  i.e. the  ensemble of  synthetic RC's, each one   derived 
from a large number  of individual RC's  belonging to  objects with a 
luminosity  falling   inside a  fixed  range (see Section 4 of PSS for
details).  The two free parameters are: the {\it velocity} core
radius $a R_{opt}$, and $(1-\beta) \ V^2_{opt}$, the halo velocity amplitude at 
$R_{opt}$.\footnote{$R_{opt}$ is  the radius   encompassing 83\% of the total 
luminosity of the galaxy. In  the  case of a (stellar) exponential thin  disk
$R_{opt}$ is 3.2  times the disk scale length $R_D$.} For the aim   of this
work, in  the region of interest,  $1/3 R_{opt}  \leq r \leq  R_{opt}$,  the
PI  circular velocity is indistinguishable  from the {\it halo } URC  of Eq. 5.
In fact, the transformation law is obtained from the known  quantities
$R_{opt}$ and $V_{opt}$, and by setting the  PI free  parameters $R_C$ and
$\rho_0$ in Eq. 4 at the values  $R_C=  a\, R_{opt}/1.45$ and $\rho_0= [  4\pi
G R_C^2 ( 1 - \frac{R_{C}}{R_{opt}}  {\rm arctan} 
\frac{R_{opt}}{R_{C}})]^{-1}  (1-\beta) V_{opt}^2$. Let us notice that    $
V_{H, PI}$ and $V_{H,URC}$ agree with each other within few percent at most, a
discrepancy that is completely negligible, in view of  the large  radial 
variations  of  $V(r)$ inside galaxies  and of  the large differences among
the  halo velocity profiles  predicted by different  mass models.    

Moreover, the  velocity profiles of Eqs. \ref{eq:pi} and \ref{eq:urc},  once
their free parameters have been properly set, can  mimic in the regions
where data are available a number of  ``theoretical'' rotation curves, not  all
necessarily consistent  with observations.\footnote{e.g. the  NFW circular
velocity  with   $r_s=R_{opt}$,   in the radial  range,   $0.3< r/r_s<1.1$ is
reproduced,  within a negligible  few percent discrepancy,    by  $V_{H, URC}
(r; a=0.33 )$ and/or  by  $V_{H, PI}(r; R_C=0.2 R_{opt})$.}

Therefore, with the  obvious caveat that  suitable data are  available,  by 
investigating the properties of the density distribution of the dark halos
around galaxies  by means of  the PI/URC profile   we  neither assume nor
reject the theoretical   $\Lambda$-CDM scenario. On the contrary,   by adopting
certain values for  the  free parameters of the chosen  halo density profile in
order  to  match the observations, we will be able to  probe  a density cusp,
or   a   core  or even  something in between.   

In this work,  we first obtain  the density  profile of DM halos around a
significant number of galaxies of different Hubble types with high-quality
kinematic data. The  intrinsic high quality of the data will allow us  to
disentangle the available kinematics in its dark and luminous contributions and
to constrain the DM halo parameters allowing us to  investigate  the existence
of a   link between their structural  properties  and  those of the ordinary
baryonic matter (i.e. stars and gas).

 The sample of data used in our analysis is presented  and discussed in Sect.
\ref{sec:sample}, while in Sect. \ref{sec:model} we describe the mass
modelling. In Sect. \ref{sec:results} we will look for a relationship between
the dark  and stellar structural  parameters.
 Finally the conclusions will be drawn in Sect.
\ref{sec:conclusions}.

\section{Mass modelling from the Rotation Curves}
\label{sec:sample}
 It is  necessary, before building a  sample of objects with  the   mass model
properly  derived from the kinematics,  to discuss the various sources of 
uncertainties relative to the task  we want to  undertake.  In fact, 
inferring the halo density {\it profile}  from a RC is complicated and it is
certain that, in spite of  applying an  effective mass modelling  method,  we get
ambiguous or incorrect results when  the observational
errors/biases/uncertainties are larger than a certain threshold.  It is easy to 
anticipate  that a main  requirement for a proper analysis  is   that the   RC's  are of 
high spatial resolution (see below) and  that they  do not  show any  significant non-circular
motion: data must allow  us to   derive,  at different radii, a suitable
estimate of  the circular velocity {\it slope}. 

In order to detail this delicate point, let us  write the    halo density
$\rho(r)$  as an explicit function of {\it  observational}  quantities,  namely
the  circular velocity,  its gradient and the  disk scale length, and  of a
{\it  derived }  quantity, the disk mass (Fall \& Efstathiou, 1980): 
\begin{equation} 
\rho(r) \simeq
G^{-1}V^2(r)/r^2(1+2 \nabla(r)) - G M_D R_{opt}^{-3} A(r/R_D) 
%\eqno(6)
\end{equation} 
where $\nabla \equiv {\rm dlog} \; V/ {\rm dlog} \; r $,  $M_D$ is the disk
mass and $A$ is a  given function (for details see the original paper). It is
easy to show that, since in the inner regions of galaxies  the two terms in the
r.h.s.  balance themselves,  even mild uncertainties in $V(r)$ trigger off very
large uncertainties in $\rho(r)$, even if $M_D$ is known. Small uncertainties
in the amplitude {\it and} in the slope of the circular velocity are an
absolute must for a proper kinematic density modelling of RC's. Moreover, a 
high spatial resolution is also  required in order to have    a sufficient 
number of  independent measurements  to beat random errors and  to  be able to
probe  the radial  constancy of the halo density in the central region,  or  to
follow the NFW profile inside  the cuspy zone. Notice also that a tight
relationship between the stellar disk scale length $R_D$ and the dark halo core
radius such as the one found in this paper cannot arise spuriously  from
uncertainties in the determination of $R_D$, as shown in Appendix A.

We select our sample of rotation curves of disk systems by  following the PSS
and Borriello and Salucci (2001)  prescriptions. The selected  RC  must satisfy
the following quality requirements: {\it 1)} data extend (at least) out to the
optical radius and to the halo core radius (when detected); {\it 2)} it is
smooth and symmetric; {\it 3)} it has small internal r.m.s.; {\it 4)} it has
high spatial resolution and a homogeneous radial coverage along both arms.  The
last  requirements can be summarised by setting a maximum uncertainty of 3\% in
the RC amplitude and 0.05 in its logarithmic derivative.

Let us stress that, when the  proper mass modelling is applied to the
kinematics of galaxies of this  suitably selected  sample, nothing   prevents
us from detecting a halo core radius of {\it zero} length or with a  value
uncorrelated to any other galaxy quantity,  if this  is how Nature has
arranged.

In Table 1 we list the sample of the galaxies selected within the above terms
that we will use later in the analysis and their relevant parameters. The
sample is not restricted to a single morphological type, but includes dwarfs,
spirals, LSBs and it considers the most reliable halo density distributions
obtained so far.
\begin{table}
% \centering
%\begin{minipage}{140mm}
\label{tab:sample}
\caption{The sample of galaxies.
(1) and (2) name and  morphological type; (3)
stellar disk scale-length   $R_D$ (kpc); (4) and (5)
core radius $R_C$ (kpc) and central density $\rho_0$
($10^{-3}$ M$_\odot$ pc$^{-3}$); (6) references:
1: Borriello \& Salucci 2001;
2: Swaters et al. 2003;
3: Weldrake et al. 2003;
4: Bottema \& Verheijen 2002;
5: de Blok \& Bosma 2002;
6: Bolatto et al. 2002;
7: de Blok et al. 2001;
8: Dutton et al. 2003;
9: Simon et al. 2003;
10: Gentile et al., 2004;
*: J. Simon, private communications}
\begin{tabular}{l|c|r|r|r|c}
\hline
Galaxy & type & $R_D$ & $R_C$ & $\rho_0$ & Ref.\\
\hspace{0.2cm} (1) & (2) & (3) & (4) & (5) & (6) \\\hline
531-G22 & Spiral & 3.3 & 5.9 & 33.4 & 1 \\
533-G4 & Spiral & 2.7 & 4.7 & 39.2 & 1 \\
563-G14 & Spiral & 2.0 & 3.5 & 60.9 & 1 \\
M-3-104 & Spiral & 1.5 & 6.3 & 37.7 & 1 \\
N755 & Spiral & 1.5 & 3.3 & 46.4 & 1 \\
U5721 & Dwarf & 0.5 & 1.1 & 123.0 & 2 \\
U8490 & Dwarf & 0.7 & 1.1 & 95.6 & 2 \\
U11707 & Dwarf & 4.3 & 17.5 & 1.2 & 2 \\
F568-V1 & LSB & 4.0 & 14.5 & 2.0 & 2 \\
N6822 & Dwarf & 0.7 & 1.7 & 33.7 & 3 \\
N3992 & Spiral & 6.5 & 23.2 & 2.5 & 4 \\
U4325 & LSB & 1.6 & 3.2 & 72.2 & 5 \\
N1560 & LSB & 1.3 & 2.1 & 29.4 & 5 \\
N3274 & LSB & 0.5 & 1.0 & 132.0 & 5 \\
N4455 & LSB & 0.7 & 2.4 & 22.6 & 5 \\
N4605 & Dwarf & 0.7$^*$ & 2.6 & 71.0 & 6 \\
F571-8 & LSB & 2.7 & 4.2 & 34.4 & 7 \\
F583-1 & LSB & 1.6 & 3.4 & 14.3 & 7 \\
N3109 & Dwarf & 1.3 & 2.4 & 24. & 8 \\
IC2574 & Dwarf & 2.2 & 6.9 & 5.9 & 8 \\
N5585 & Spiral & 1.6 & 2.2 & 42.5 & 8 \\
N2403 & Spiral & 2.1 & 4.8 & 17.7 & 8 \\
N2976 & Dwarf & 0.7$^*$ & 1.1 & 100.0 & 9 \\
116-G12 & Spiral & 1.7 & 2.2 & 53.2 & 10 \\
79-G14  & Spiral & 4.1 & 5.5 & 23.6 & 10 \\
\end{tabular}
%\end{minipage}
\end{table}

The distances of the galaxies have been taken directly in the original papers.
They are computed by means of local calibrators or assuming Hubble flow after
correction for Galactic rotation and Virgocentric flow,  with $H_0$=75
km/sec/Mpc. The uncertainties on the distances can be estimated about 0.30 mag,
which implies an uncertainty on any galaxy  logarithmic linear size of 0.12
dex. 

Finally, let us point out that  1) in literature there are about 10 other
objects  for  which the  kinematics,   although it fails our selection
requirements, still   provides a   solid evidence for the presence of
a core radius of which it is however unable to   constrain the  size (e.g.
Salucci et al, 2003); 2) we failed to find  objects of high/average  quality
kinematics  that {\it clearly  prefer}  the  NFW profile.  

%%%%%%%%%%%%%%%%%%%%%%%%%%%%%%%%%%%%%%%%%%%%%%%%%%%%%%%%%%%%%%%%%%
%%%%%%%%%%%%%%%%%%%%%%%%%%%%%%%%%%%%%%%%%%%%%%%%%%%%%%%%%%%%%%%%%%
\section{ Mass models}
\label{sec:model}
 
Let us  detail the mass model, and Freeman disk+ HI disk + pseudo-isothermal
dark halo  used  to best fit $V(r)$. 

\subsection{Baryonic Disks}

It is usually assumed that light traces the stellar mass via a radially
constant mass--to--light ratio, so that the stellar disk contribution is 
given by ($y=r/R_{opt}$):
\begin{equation} 
V^2_{D}(y)= 14.7 ~ V_{D,opt}^2~y^2~(I_0K_0-I_1K_1)|_{1.6y}
%\eqno(7) 
\end{equation} 
where $V_{D,opt}\equiv V_D(R_{opt})$.
The circular velocity of the gas has also been included and derived from the
surface density of the HI disk. Gaseous helium and metals have been taken into
account by a simple rescaling factor of the surface density of 1.33. The
contribution of the gaseous disk  to the total rotation speed is found to be
marginal. Only in the analysis of Borriello \& Salucci (2001) and of Bolatto et
al. (2002) the gas has not been included. However, both the authors note that
its contribution in the explored region would not modify any of their results.

\subsection{Dark Halos}

We choose the  pseudo--isothermal halo profile PI,  adopted  to model  20/25 of
the sample RC's,  as  the reference halo profile of  this work
\begin{equation}
\rho_{PI}(r) = \rho_0 \frac{R_C^2}{r^2 + R_C^2}
%\eqno(8)
\label{eq:rho_isoth}
\end{equation}
where $\rho_0\equiv \rho_{PI}(0)$ and $R_C$ is the PI  core radius. We  link  
the core radius of  this profile with that of the   URC  profile,   adopted to 
model  the Universal Rotation  Curve, for the remaining 5 RC's of our sample 
(those from Borriello \& Salucci, 2001)
and the sample of ellipticals we consider in the conclusions,  by recalling
that (see previous section) the two  profiles almost coincide when   $R_C = 
a  R_{opt}/ 1.45 $ (see also, for  some  examples, Gentile et al. 2004).    

The $R_C$ of NGC 4605 in Bolatto et al. (2002) was fitted with in the original
paper  power-law density $\rho \propto r^{-0.65}$. To homogenise it with the
analysis on all the other galaxies we have fitted their best fit profile in
terms of Eq. \ref{eq:rho_isoth} and derived $R_C$=2.6 kpc.

We remind the reader that the free parameters of our fitting procedure are 
three, namely: $V_D^2(R_{opt})$, $\rho_0$ (or $\beta$ for the URC RCs) and 
$R_C$. Finally, let us point out that, for most of the mass models in  our 
Sample a maximun disk {\it is not assumed}, but the disk {\it is found} to be
dominant at $R_D$ , as result of  a   $\chi^2$ fitting.

%%%%%%%%%%%%%%%%%%%%%%%%%%%%%%%%%%%%%%%%%%%%%%%%%%%%%%%%%%%%%%%%%%%%
\section{Results}
\label{sec:results}
A  relationship  between ``halo core radii" and galaxy  luminosities was 
hinted  in    Persic and Salucci (1988) and in an  early (unrefereed)  analysis
of galaxy structure  by Kormendy (1990). Subsequently,   PSS  found that the
synthetic RC's  define an Universal Rotation Curve whose dark halo component 
has a  core radius related to disk scale lengths and I-band magnitudes:   $
R_C= 3.5  R_D~10^{-(0.2 \pm 0.1) \ (M_I-21.9)/2.5 }$. This result   can be
considered as a serious   indication  that the  DM {\it distribution}  
relates  to  the properties of the luminous matter. It is, however  only
indicative in that: 1)  the  RC's coaddition procedure and  the relative
building of the URC  was  optimised  to properly  derive  the radial behaviour 
of the  dark-to-luminous {\it mass ratio} rather than the inner slope of the
dark halo density.  2) The relationship we are  looking  for must be primarily
searched for in a  sample of {\it individual} objects, in order to relate  the
relevant  quantities in each individual object.

In this Section we  investigate the properties of this ``new entry" in galactic
modelling, the DM core length $R_C$, i.e.  the scale length inside which the
dark matter density varies very mildly, and outside which, it starts to decline
as a power law. Is $R_C$ only a consequence of the fact that a cuspy
distribution is a poor representation of actual DM density, or it is a real
physical quantity related to the dynamical properties of the galactic systems?
Moreover, could core radii arise from observational errors, as recently
suggested (Swaters et al., 2003)? 

The  values of core radii  for the objects in  our sample, shown in Table 1,
help to answer to these questions, revealing a tight correlation between $R_C$
and $R_D$ that we show in Fig. 1. This  is well fitted by:
\begin{equation}
\log R_C = (1.05\pm 0.11) \; \log R_D + (0.33\pm 0.04)
%\eqno(9)
\label{eq:Rc_Rd}
\end{equation}
\[ corr = 0.90, \; \; r.m.s. = 0.16 \; dex  \]
with $corr$ being the correlation number and {\it r.m.s.}
the least square scatter. The  relation clearly exists at very high statistical
confidence, in addition, LSBs  and dwarfs   follow the spirals' distribution
around the mean line  with no segregation.
\begin{figure}
%\vspace{1cm}
\vspace{0.5cm}
\includegraphics[width=95mm]{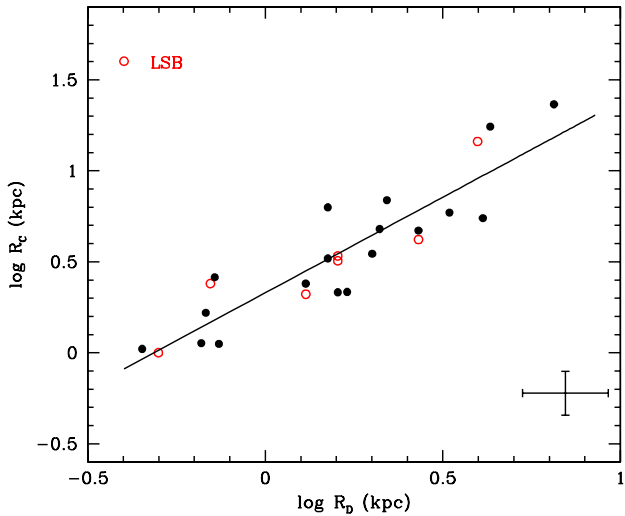}
\label{fog:Rc_Rd}
\vspace{-0.5cm}
\caption{Core radius $R_C$ as a function of the disk scale
length $R_D$. Open and filled circles refer to LSB and HSB galaxies, 
respectively. The solid line  is the least square fit.} 
\end{figure}
\begin{figure}
\vspace{0.5cm}
\includegraphics[width=95mm]{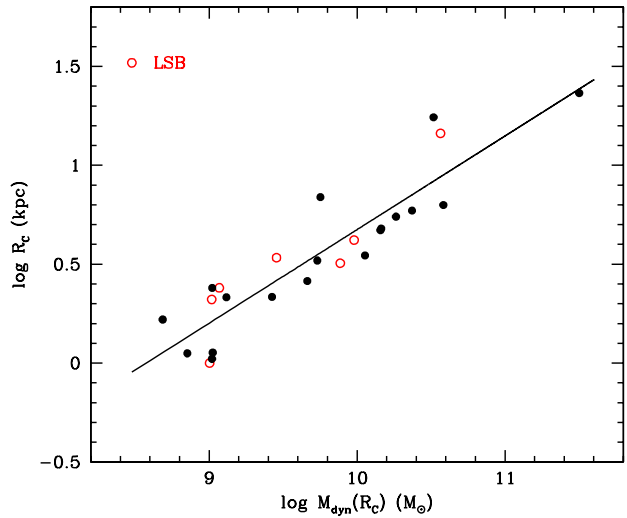}
\vspace{-0.5cm}
\caption{Core radius $R_C$ as a function of the dynamical mass inside
$R_C$.  Open and filled circles refer to LSB and HSB galaxies, 
respectively. The solid line is the least square fit.}
\label{fig:Rc_Mdyn}
\end{figure}

As stressed above,  $R_D$ is derived by mere photometric measurements, while
$R_C$ is obtained from the kinematics of the regions dominated by the dark
matter.  A spurious link between these two quantities cannot exist in that they
are measured in separate ways, and the tight correlation between $R_C$ and
$R_D$ cannot be induced by errors in the determination of $R_D$ (see Appendix
A). Then, if: 1) the  cored profile did not describe the actual dark matter
density in galaxies or 2) the kinematics had serious observational
errors/biases or 3) the dark/luminous decomposition was wrongly made, it is
very difficult to imagine how such a tight relationship could have emerged.

We find also a second tight  correlation  involving the core
radius and the dynamical galaxy mass  inside it,
$M_{dyn}(R_C) \equiv G^{-1} V^2(R_C) R_C$,  that we plot in
Fig. \ref{fig:Rc_Mdyn}:
\begin{equation}
\log R_C = (0.47\pm 0.05) \log \left( \frac{M_{dyn}(R_C)}{10^{11}M_\odot} 
\right) + 
(1.15\pm 0.07)
\label{eq:Rc_Mdyn}
\end{equation}
\[ corr=0.91, \;\; r.m.s. = 0.15 \; dex.  \]
This  is an additional  link between an observed quantity, $V(r)$  and the core
size $R_C$ and there is  no conceivable reason that would explain such a very 
tight relation in terms of observational or modelling biases, while it is very
convincing that it simply indicates that the bigger the galaxy, the bigger the
core radius.  Notice that the slope of 0.47, very different from 1 in Eq.
\ref{eq:Rc_Mdyn} --  that relates two variables each one linearly dependent on 
galactic distance -- implies that  uncertainties on the latter affect very
little {\it both} relationships in  Eq. \ref{eq:Rc_Rd}  and  in Eq
\ref{eq:Rc_Mdyn}.

\begin{figure}
\vspace{1cm}
\includegraphics[width=95mm]{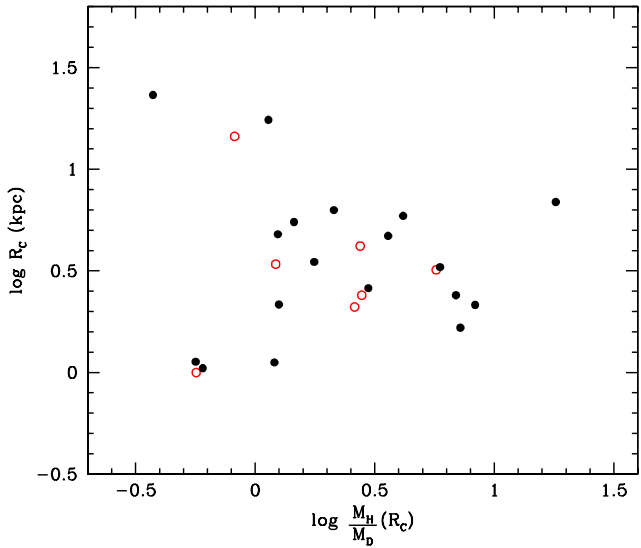}
\caption{Core radius $R_C$ as a function of the disk to halo mass ratio at
$R_C$.  Open and filled circles refer to LSB and HSB galaxies,  respectively.}
\label{fig:Rc_MhsuMd}
\end{figure}

Finally, in the case of incorrect  mass  decomposition of RC's, we would 
expect  a particular  spurious  correlation between the core radius and  the
dark-to-luminous mass fraction (MF) at  $R_C$,  in that both concur and compete 
to shape the RC profile (see Persic and Salucci 1990). In Fig.
\ref{fig:Rc_MhsuMd} we plot $R_C$ as a function of this fraction at  $r=R_C$:
$M_H(R_C)/M_D(R_C)$. The absence of a tight  correlation is evident, and
demonstrates that the  $R_C$ values  are not affected by gross misfits in the
mass decomposition and, more importantly, that the size of the  core radii does
not   seem to be  primarily dependent on the  fractional amount of DM in this
region of galaxies. 

Let us further  notice that the above  lack of a  tight correlation cannot be
related  to  the existence, at $3 R_D$,  of the  trend between  similar
quantities, as implied by the results in   PSS. By recalling that  such
relationship is found only  at   fixed multiples of  disk scale lengths, we 
point out that   the scatter of  0.16 dex in  the $\log R_C $ vs $\log R_D $ 
relationship, though statistically  small with respect to the length-scale
range,  is however  large enough  to imply that $R_C$   corresponds  to  very
different galacto-centric  distances in  different galaxies, {\it i.e.} it
corresponds  to  distances varying from  $1.5  R_D$    to $ 3.1 R_D$.  As a 
result, the  DM fractions computed  at   $R_C$'s    are actually  computed at
a  very different radii in different objects; and we do not expect to find
there  the correlations we found at each  fixed multiple of  $R_D$.

%%%%%%%%%%%%%%%%%%%%%%%%%%%%%%%%%%%%%%%%%%%%%%%%%%%%%%%%%%%%%
\section{Conclusions}
\label{sec:conclusions}
There is a mounting evidence that the  kinematics of disk galaxies can be
explained in terms of the standard  disk + halo components {\it only } if the 
density distribution of the latter decreases with radius  very slowly from the
center out to the radius  $R_C$  inside which, instead, the density of CDM
halos is predicted to fall  as a power law with an exponent between -1 and -2.
Here we have investigated this ``observational" dark matter scale $R_C$  that is
absent in the current theory of galaxy formation. We found that it correlates
tightly with the exponential thin disk scale length $R_D$  and with the amount
of gravitational mass that it encompasses $M_{dyn}(R_C)$. The high values for
the correlation number, the smallness of the scatter around the relations and
especially the fact that the pairs $R_C$ and $R_D$, as well as $R_C$ and
$M_{dyn}(R_C)$, are measured/derived independently from each others, ensure
that we are dealing with {\it real and physical} relationships among physical
quantities. 

Three different levels of consequences follow. First, the claims according to
which core radii arise as a consequence of 1) serious observational errors, 2)
peculiar or biased kinematics, 3) wrong mass modelling that, until today, have
been  { \it by-passed}  by considering a number of test-cases, can be now 
ruled out. This, on the basis that  human/observational error hardly  relate to
intrinsic properties of galaxies. As an example the  suggestion that observers
have badly missed the galaxy center and doing so artificially created a core,
cannot explain the additional evidence that this core  is found to correlate
with the galaxy disk scale length within the extraordinarily small scatter of
0.16 dex.

The second level realises that the core radius,  the quantity that defines the
unexplained  feature of the distribution of the dark matter, correlates with
the main quantity controlling the distribution of the luminous matter as well
as the total mass within this  radius. This indicates that the DM density in
galaxies has been shaped by a dark--to--luminous matter coupling, a  very
challenging task  if the former is collisionless. However, let us recall that,
at least in dwarf systems,  according to  Navarro et al. (1996a), large mass
outflows  from  galaxies, arising from supernovae explosions,  can modify   an
initial  NFW profile into a cored one,   for which  $R_C \propto
(M_D/R_D)^{0.5}$. From the results of the mass modelling given in PSS, we find
$M_D \propto R_D^3$, in excellent agreement with relationship (9). In any case, 
our  result points  toward some   new physics of which it  will be  a crucial 
benchmark.

The third level is  the realization that the core radius vs disk scale length
relationship is  Hubble type free, at least   within  a band of   0.16 dex. 
For the objects in our sample   normal/dwarf spirals and  LSB lay on the same 
$\log R_C$  vs $\log R_D$ relationship. Moreover, pioneering analysis on the DM
distribution in ellipticals (Gerhard et al. 2001) finds  halo core radii of
sizes proportional to the half light radii (that correspond to 1.67 $R_D$):
by analysing the mass model parameters of 21 Giant Ellipticals (shown 
in Fig 18 Gerhard et al. (2001), and made available by the authors) we find:
\begin{equation}
\log R_C= (1.1 \pm 0.2 )\  \log R_D +(0.1 \pm 0.4)  
\end{equation}
in very good agreement with Eq. (9).
We caution that  the sample of this work is quite  limited to fully
investigate   the morphological dependence  of DM properties and that  in
Gerhard et al.(2001)   the data quality  is barely sufficient for the 
intrinsically complicated  mass  modelling. Nevertheless,  the emerging
picture  is truly impressive,  especially  when we consider that core radii of
galaxies of different Hubble types  do not  correlate  with  galaxy luminosity:
e.g.  a  similar value for  $R_C$  is   found in a  $10^{11} L_{\odot,r}$
spiral,  in a  $10^{10} L_{\odot,r}$ LSB and  in a $10^{12} L_{\odot,r}$ giant 
elliptical. The core radius  seems to  uniquely relate to the  exponential 
rate with  which  the stellar surface density decreases with
radius.   

The next  step of this line of research post-$\Lambda$-CDM crisis  would be to
thoroughly investigate this  DM distribution property in galaxies of
different size and Hubble Types  (Salucci et al., 2004) and  the brightest
galaxies  because of the slightly non-linear  relationship found between  $R_C$
and the stellar scale length. 

\section*{Acknowledgements} P.S. thanks 
I. Yegorova for help in the presentation.

%\appendix 
\section{Appendix A}
For  sake of completeness, we investigate how much   observational   errors in
$R_D$ could  bias the derivation of the halo density profile and, in
particular, the determination of the core radius size  $R_C$.  We  checked this
by means of the 4 high-resolution RC's of spirals with an exponential  thin
disk stellar distribution  published in   Gentile et al. (2004). 
\footnote{Two of these RC's  meet  the requirements in Sect. 2; 
the other two  are, however, of
a quality  sufficient for the  aim of  this Appendix: notice below that  the
effect  investigated here   leads to intrinsic    variations of halo core
radii   much  larger than  those that arise from  the  observational
uncertainties of the RC's.} First, we  adopted   for  $R_D$ the photometric
values, we best fit the RC's  and derived   the DM halo core radii   $R_C$'s.
We then let the   $R_D$'s  to  vary  by an  amount up to  $ \pm 30$\%, 
assuming in this way  that the photometric values were  wrong by this amount.
Finally, we  best-fit again   the RC's with these new    $R_D$ values  and
derived  the  corresponding  new $R_C$ values.

As a result,  we find  that values of $R_D$ even mildly different from the
photometric  ones could sometime lead to unacceptably high  $\chi^2$'s for any
value of the free parameters (for example, for the NGC1090 galaxy, 
$\frac{R_D(wrong)}{R_D(true)}$=0.70 leads to a reduced $\chi^2=2.45$, about
five times greater than the one obtained for $R_D(true)$, which is
$\chi^2=0.53$). This is in agreement 
with the fact that for  $r<R_D$, rotation curves are largely  dominated by the
disk that inprints on it   its   {\it true}  disk scale-length value (see
Salucci et al., 2000; Ratnam \& Salucci, 2000).   
 
\begin{figure}
\centerline{
\resizebox{0.9\columnwidth}{!}
{\includegraphics*[width=75mm]{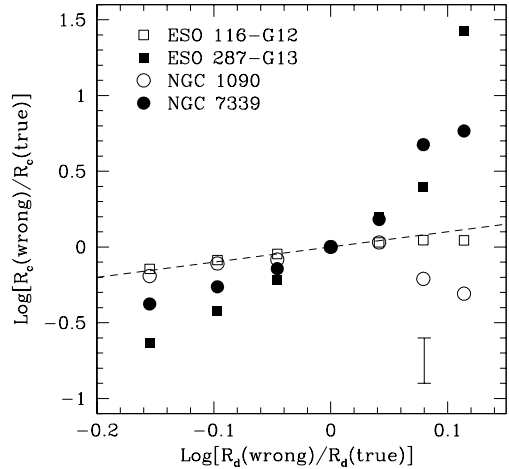}}}
%\vspace{-1.3cm}
\caption{Changes of values of  core radius as a function of the disk scale 
length fractional errors.}
\label{fig:errors}
\end{figure}

The study of the fractional  error in the estimate of core radius 
$\frac{R_C(R_D(wrong))}{R_C(R_D(true))}$ as a function  of  the (assumed) error
in the disk scale length $\frac{R_D(wrong)}{R_D(true)}$   is shown in Fig.
\ref{fig:errors}.  
 
We realize that  an  observational error  in $R_D$   may   
trigger  an   error   $\Delta R_C$   in the  inferred value  of $R_C$ we 
get by fitting the same RC with the wrong value of $R_D$:  $\Delta 
R_c\equiv  \log ({R_C(R_D(wrong))}/{R_C(R_D(true)))} $. We have 2 
possibilities:

{\it Large} errors ($>10$\%):  $ \Delta R_C$ can be large. However:  i)   in
different objects the same (large) error in $R_D$  induces  a very different 
$\Delta R_C$:   at a fixed $\log (R_D(wrong)/R_D(true))$  the former   spans
values that are different among themselves also  by 0.5-1 dex  (see Figure 4).
This  is  inconsistent with  the observed scatter of 0.15 dex of  the  $\log
R_C$  vs  $\log R_D$  relationship;  ii) the slopes of the spurious
relationship that gets  created in  each object  are  completely different
from the observed one of +1 (see  Figure 4).

 {\it Small} ($<10$\%) errors: in this most likely case (in our  sample the
typical uncertainty is 5-10\%\footnote{ The uncertainty on $R_D$ was given in
the original paper only in some cases; in the other cases it was estimated by
us.})   $\Delta R_C$  turns out to be related with disk scale lengths (see
Figure 4), but only at the  level  of   $<0.1$ dex, and therefore the
photometric errors cannot  possibly   create  the  observed range of 1.2 dex 
shown by   relationship (9).

 We conclude the investigation by stressing that we can consider,
for the aims of this paper, the (kinematic) measure of $R_C$ and the
(photometric) measure of $R_D$ as  independent.

\end{document}